%% file: phep-2025-298_2ha.tex
\documentclass[a4paper,manyauthors,nocleardouble,COMPASS]{cernphprep}

\RequirePackage[T1]{fontenc}

\RequirePackage{graphicx}
\RequirePackage{newtxtext}
\RequirePackage{newtxmath}
\RequirePackage{flushend}
\RequirePackage[numbers,sort&compress]{natbib}

\usepackage{etex}
\usepackage{amsmath,amssymb,epsfig,fleqn,url,color,here,graphics,graphicx,rotating,multirow,wrapfig}
\usepackage{pbox}
\usepackage{dcolumn}
\usepackage{bigstrut}
\usepackage{soul}
\usepackage{lineno, xcolor}
\usepackage{balance}
\usepackage{lastpage}

\usepackage[section]{placeins}
\usepackage{orcidlink}
\usepackage{tikz}
\usepackage{hyperref}
\hypersetup{
    colorlinks,
    citecolor=blue,
    filecolor=blue,
    linkcolor=blue,
    urlcolor=blue
}
\newcommand{\xbj}{x}                             
\newcommand{\Qs}{Q^2}                             
\newcommand{\zh}{z}                              
\newcommand{\qvec}{\mathbf{q}}                   
\newcommand{\lvect}{\mathbf{l}}                   
\newcommand{\pvec}{\mathbf{p}}                   
\newcommand{\Rvec}{\mathbf{R}}                   
\newcommand{\Svec}{\mathbf{S}}                   

\newcommand{\Mhh}{M_{\rm{hh}}}                   
\newcommand{\Mhhsq}{M_{\rm{hh}}^2}               
\newcommand{\phiR}{\phi_{\rm R}}                 
\newcommand{\phiS}{\phi_{\rm S}}                 
\newcommand{\phiRS}{\phi_{\rm RS}}               

\newcommand{\PT}{P_{\rm T}}                      
\newcommand{\Dnn}{D_{\rm nn}(y)}                 

\newcommand{\iFF}{H_1^{\sphericalangle}}  

\newcommand{\gvc}{GeV/$c$}
\newcommand{\gvcs}{(GeV/$c$)$^2$}
\newcommand{\gvcw}{GeV/$c^2$}
\newcommand{\dd}{\mathop{}\mathopen{}\mathrm{d}}

\setstcolor{red}
\makeatletter
\g@addto@macro\bfseries{\boldmath}
\makeatother
\begin{document}
\begin{titlepage}
\PHnumber{2025-298}
\PHdate{23 Decembre 2025}

\title{Dihadron Transverse-Spin Asymmetries in Muon-Deuteron Deep-Inelastic Scattering}
\begin{abstract}
In 2022, the COMPASS collaboration performed semi-inclusive measurements of
deep-inelastic muon-scattering on a transversely polarised deuteron ($^{6}$LiD)
target.  From these data, transverse-spin–dependent dihadron asymmetries are
extracted using pairs of oppositely charged hadrons.  These asymmetries are
directly sensitive to the quark transversity distributions and provide an
independent handle on these fundamental quantities with respect to the Collins
asymmetries measured in single-hadron production.  The present results
significantly improve upon the previous COMPASS deuteron measurements, which
were the only available deuteron data worldwide, and reach a statistical
precision comparable to that of the existing proton results from COMPASS.  A
small but nonzero asymmetry is observed at large Bjorken-$\xbj$, consistent with
theoretical expectations.  A point-by-point extraction of the valence-quark
transversity distributions yields, in particular, a substantially improved
determination of the $d$-quark transversity.  These measurements represent a
major step towards a complete flavour mapping of the transverse-spin structure
of the nucleon.
\end{abstract}

\Collaboration{The COMPASS Collaboration}
\ShortAuthor{The COMPASS Collaboration}
\vfill
\Submitted{(to be submitted to Phys. Rev. Letters)}
\end{titlepage}

\date{\today}
\titlepage

{
\pagestyle{empty}
\input{Authors_list_2025.11.22_PRL_D_SIDIS_2h_TSAs}

\clearpage
}
\clearpage

\setcounter{page}{1}

%
%

\section{Introduction}
In quantum chromodynamics (QCD), a fast-moving nucleon (proton or neutron) is a
composite, dynamic system of quarks and gluons (partons) exhibiting both
longitudinal and transverse motion.  The collinear parton distribution functions
(PDFs) describe the longitudinal momentum fraction $\xbj$ carried by partons
inside the nucleon~\cite{Feynman:1972photon,Collins:2011zzd} and form the
cornerstone of QCD phenomenology, providing the foundation for global analyses
and precision predictions in high-energy processes.  The
transverse-momentum-dependent (TMD) PDFs extend this framework by incorporating
the partonic transverse momentum $\textbf{k}_T$, thereby offering access to the
three-dimensional momentum structure of the
nucleon~\cite{Kotzinian:1994dv,Mulders:1995dh,Bacchetta:2006tn}.

At leading twist, the full quark spin and momentum structure is described by
eight TMD PDFs for each quark flavour, encompassing all possible correlations
between quark spin, transverse momentum, and nucleon spin.  Among them, the
spin-averaged $f_1(\xbj,\textbf{k}_T^{2})$, the helicity
$g_1(\xbj,\textbf{k}_T^{2})$, and the transversity $h_1(\xbj,\textbf{k}_T^{2})$
distributions have collinear counterparts: upon integration over $\textbf{k}_T$,
they reduce to $f_1(\xbj)$, $g_1(\xbj)$, and $h_1(\xbj)$.  The remaining five
TMDs vanish upon $\textbf{k}_T$-integration.

The collinear transversity distribution $h_1(\xbj)$ describe transversely
polarized quarks inside a transversely polarized
nucleon~\cite{Ralston:1979ys,Artru:1989zv,Jaffe:1991ra,Barone:2001sp,Collins:2011zzd}.
It represents the difference in the probability for finding a quark with spin
aligned versus anti-aligned with respect to the transverse spin of the
nucleon. It is directly related to the nucleon tensor charge, currently being
evaluated in lattice QCD~\cite{Alexandrou:2018eet,Alexandrou:2024ozj}.  Being a
chiral-odd function, $h_1$ can be measured only through its coupling to another
chiral-odd partner.  This property distinguishes it from chiral-even $f_1$ and
$g_1$ and prevents its determination in inclusive deep-inelastic scattering
(DIS).

At variance, the TMD transversity distribution $h_1(\xbj, \textbf{k}_T^{2})$ can
be accessed in single hadron production in semi-inclusive DIS (SIDIS) with
transversely polarized targets through its convolution with the chiral-odd
Collins fragmentation function $H_1^{\perp}$~\cite{Collins:1992kk}. In the
Drell–Yan process it couples with another transversity distribution or with the
chiral-odd Boer–Mulders TMD PDF~\cite{Boer:1997nt,Arnold:2008kf}.  In TMD
factorization framework the observable asymmetries involve convolutions in
transverse momentum.  As a consequence, the determination of the TMD PDFs and
FFs requires assumptions or parametrizations of their transverse-momentum
shapes~\cite{Collins:1981uw,Collins:2011zzd}. Alternatively, one can employ
various weighting techniques to access transverse momentum moments of the PDFs
and
FFs~\cite{Kotzinian:1995cz,Kotzinian:1997wt,Boer:2011xd,Aghasyan:2014zma,COMPASS:2018ofp}.

In practice, SIDIS
measurements~\cite{HERMES:2004mhh,COMPASS:2006mkl,COMPASS:2012ozz,COMPASS:2014bze,HERMES:2020ifk,COMPASS:2023vhr}
offer significantly higher statistical precision than Drell--Yan
data~\cite{COMPASS:2017jbv,COMPASS:2023vqt} and enable a more direct flavour
decomposition through the combined use of proton and deuteron (or neutron)
targets and charge-separated hadron samples. For this reason, most of the
present knowledge on the transversity distribution $h_1$ comes from global
analyses that combine SIDIS and $e^+e^-$ data, the latter providing essential
information on the chiral-odd fragmentation functions (FFs) required for the
extraction.

Despite the fact that the collinear transversity PDF $h_1(\xbj)$ cannot be
accessed in inclusive DIS, it can be measured in semi-inclusive production of
oppositely charged hadron-pairs, where it couples to the polarized dihadron
interference fragmentation function
$\iFF$~\cite{Collins:1994ax,Jaffe:1997hf,Bianconi:1999cd}. This approach is
particularly important and unique because it allows for a direct extraction of
$h_1(\xbj)$ without requiring transverse-momentum–dependent factorization,
thereby avoiding complications related to transverse-momentum convolutions and
soft-factor effects. The same dihadron framework also applies to polarized
proton-proton collisions, providing an independent access to
$h_1(\xbj)$~\cite{Radici:2016lam}.

The dihadron fragmentation function $\iFF$, which couples to transversity in
collinear factorization, can originate from different mechanisms. One arises
from the interference between quantum amplitudes associated with two competing
channels for hadron-pair production, such as the direct production of the pair
versus its production through a resonance
decay~\cite{Collins:1994ax,Bacchetta:IFF_rho0}, or the production of the pair
from the decays of two different resonances~\cite{Jaffe:1997hf}. Another
mechanism is provided by the string+$^3P_0$ model, in which the string
connecting, e.g. the scattered quark and the target remnant in a DIS event,
breaks via the tunneling of quark–antiquark pairs with correlated spins and
transverse momenta in the relative $^3P_0$
state~\cite{Artru:2010st,Kerbizi:2018qpp}.  As in the case of the Collins FF,
the $H_1^{\sphericalangle}$ has been extracted in global analyses combining
$e^+e^-$ annihilation data from the Belle Collaboration~\cite{Belle:2011cur}
with SIDIS data, where it is determined simultaneously with the transversity
distribution~\cite{Courtoy:2012ry,Radici:2015mwa,Radici:2018iag}.

The dihadron TSAs have been measured to be nonzero by
HERMES~\cite{HERMES:2008mcr} and by
COMPASS~\cite{COMPASS:2012bfl,COMPASS:2014ysd,COMPASS:2023cgk} on transversely
polarised proton targets. For flavour separation, measurements with neutron or
deuteron targets are required. Prior to this work, the only such result was the
COMPASS measurement with a transversely polarised deuteron target from the
2002–2004 data set~\cite{COMPASS:2012bfl}, where the asymmetries were found to
be compatible with zero within uncertainties.

This work presents new COMPASS results on dihadron TSAs, extracted from the 2022
data set collected with a transversely polarised deuteron target. It constitutes
a natural continuation of the single-hadron transverse spin asymmetry analysis
recently published in Ref.~\cite{COMPASS:2023vhr}, now extended to the dihadron
channel.

\section{Formalism}
\begin{figure}[h!]
\centering
\includegraphics[width=0.50\textwidth]{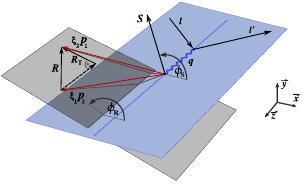}
\caption{Dihadron production coordinate frame and definition of the relevant vectors and azimuthal angles.}
\label{fig:frame}
\end{figure}

The coordinate frame for dihadron production in SIDIS and the definitions of the
relevant angles are provided in Fig.~\ref{fig:frame}.  The relevant momenta are
the incoming lepton momentum $\lvect$, the virtual-photon momentum $\qvec$, and
the relative hadron momentum $\Rvec = \xi_2\pvec_1 - \xi_1\pvec_2$, where
$\pvec_1$ and $\pvec_2$ are the momenta of the two selected final-state hadrons,
and $\xi_i = z_i/(z_1 + z_2)$ with $z_1$ and $z_2$ the fractions of the
virtual-photon energy carried by hadron~1 (positive) and hadron~2 (negative),
respectively. The azimuthal angle $\phi_S$ of the spin of the nucleon is defined
between the lepton scattering plane (spanned by $\lvect$ and $\lvect'$) and the
transverse-spin component $\Svec_T$, while $\phi_R$ denotes the angle between
the lepton scattering plane and the plane containing $\qvec$ and the transverse
component of the relative momentum $\Rvec_{\rm{T}}$.

At leading twist and after integration over the total transverse momentum of the
hadron pair, the differential cross section for dihadron production on a
transversely polarized nucleon can then be written
as~\cite{Bacchetta:IFF_rho0,COMPASS:2012bfl}
\begin{align}
\frac{\dd^7\sigma}{\dd\cos\theta\dd \Mhh\dd\phiR\dd \zh\dd \xbj\dd y\dd \phiS} = \frac{\alpha^2}{2\pi \Qs y} \times\\
&\hspace{-6cm} \left( (1-y+\frac{y^2}{2})F_{\rm{UU}} \right.
\left. + S_{T}(1-y) \sin{\theta}F_{\rm{UT}}^{\sin{\phiRS}}\sin{\phiRS} \right). \nonumber
\label{eq:xsection}
\end{align}
Here, $\alpha$ is the fine-structure constant. The variable $y$ denotes the
fraction of the lepton energy (in the laboratory frame) transferred to the
virtual photon, and $Q^{2}$ is the negative squared four-momentum transfer. The
variables $\zh=z_1 + z_2$, $\Mhh$, and $\theta$ correspond to the fraction of
virtual-photon energy carried by the hadron pair, its invariant mass, and the
polar angle of the positive hadron with respect to the two-hadron boost axis in
the pair rest frame.  Here, $\phiRS = \phi_{\rm{R}} - \phi_{\rm{S}^{\prime}} =
\phi_{\rm{R}} + \phi_{\rm{S}} - \pi$, where $\phi_{\rm{S}^{\prime}}$ is the
azimuthal angle of the spin vector of the fragmenting
quark~\cite{COMPASS:2006mkl}.

The structure function $F_{UU}$ represents the spin-averaged contribution to the
dihadron production cross section, corresponding to the case of an unpolarized
beam and an unpolarized target. At leading twist, it is given by
\begin{equation}
F_{\rm{UU}} = \sum_q e_q^2~f_1^q(\xbj)~D_{1,q}(\zh,\Mhhsq,\cos\theta).
\label{eq:FUU}
\end{equation}
The sum runs over the quark and antiquark flavours $q$ with charge $e_q$, and
$D_{1,q}(\zh,\Mhhsq,\cos\theta)$ denotes the unpolarized quark dihadron
fragmentation function (FF).

The spin-dependent term $F_{UT}^{\sin\phiRS}$ is proportional to the product of
the transversity PDF $h_1^q(\xbj)$ and the dihadron FF of transversely polarized
quarks $H_{1,q}^{\sphericalangle}(\zh, \Mhhsq, \cos\theta)$:
\begin{equation}
\hspace{-0.9cm}F_{\rm{UT}}^{\sin{\phiRS}}=\frac{|\textbf{p}_1-\textbf{p}_2|}{2\Mhh}\sum_q e_q^2~h_1^q(\xbj)~H_{1,q}^{\sphericalangle}(\zh,\Mhhsq,\cos\theta) \nonumber
\label{eq:FUT}
\end{equation}
The dihadron TSA $A_{\rm{UT}}^{\sin\phiRS}$ is given by the ratio of the two structure functions:
\begin{equation}
\hspace{-0.9cm}A_{\rm UT}^{\sin\phiRS} = \frac{|\bf{p}_1-\bf{p}_2|}{2\Mhh} \frac{\sum_q\,e_q^2~h_1^q(\xbj)~H_{1,q}^{\sphericalangle}(\zh,\Mhhsq,\cos\theta)}{\sum_q e_q^2~f_1^q(\xbj)~D_{1,q}(\zh,\Mhhsq,\cos\theta)}.
\end{equation}
The dihadron TSA is extracted from the number of hadron pairs
\begin{eqnarray}
  \label{eq:Nhh}
N_{\rm{hh}}(\xbj,y,\zh,\Mhhsq,\cos\theta,\phiRS) \propto& \\
&\hspace{-3cm}\sigma_{\rm{UU}} \big( 1 + f(\xbj,y) P_{\rm{T}} D_{\rm{nn}}(y) \sin\theta A_{\rm{UT}}^{\sin\phiRS} \sin\phiRS \big), \nonumber
\end{eqnarray}
where $f(\xbj, y)$ is the target-polarisation dilution factor accounting for the
fraction of polarizable nucleons in the target material, $P_{\rm{T}}$ is the
transverse target polarisation, and $D_{\rm{nn}}(y)$ is the transverse-spin
transfer coefficient.

\section{Experimental Data and Event Selection}

The present analysis closely follows the procedures established in previous
COMPASS dihadron studies~\cite{COMPASS:2012bfl,COMPASS:2014ysd}, including event
selection criteria, kinematic cuts, and the extraction method for azimuthal
asymmetries.  The data analyzed in this work are the same as those used for the
Collins and Sivers asymmetry measurements reported in
Ref.~\cite{COMPASS:2023vhr}, where the experimental apparatus, data taking, and
event selection are described.  The data were collected by the COMPASS
experiment~\cite{COMPASS:2007rjf} at CERN during the 2022 run, using a 160
\gvc\, naturally polarized $\mu^+$ beam scattering off a transversely polarized
deuteron ($^6$LiD) target. The experimental setup included a two-stage
spectrometer equipped with tracking detectors, calorimetry, and particle
identification systems to ensure precise reconstruction of the scattered muon
and produced hadrons.

The analysis is based on events that fulfill deep-inelastic scattering
conditions: $\Qs > 1$ \gvcs, $0.1 < y < 0.9$ and hadronic invariant mass $W > 5$
\gvcw. For dihadron studies, at least two oppositely charged hadrons originating
from the interaction vertex are required. These hadrons must have $z_{1,2} >
0.1$ and $\zh = z_1 + z_2 < 0.9$ to suppress exclusive contributions.  The
selection $\Mhh>$ 0.3 \gvcw\, is adopted to avoid the low-mass threshold region.

The target consisted of three cylindrical cells, 30, 60, and 30~cm in length and
1.5~cm in radius, which were polarised in alternating directions to minimise
acceptance effects. The target polarization was reversed in the middle of each
of the ten data-taking periods of the 2022 run to minimize systematic
effects. Data from the two polarization orientations were then combined to
extract the transverse-spin asymmetries.

The TSAs are extracted by fitting the distribution of hadron-pair yields
[Eq.~(\ref{eq:Nhh})] as a function of $\phiRS$ using an unbinned
maximum-likelihood method. The extracted observable is $\langle
A_{UT}^{\sin\phi_{RS}}\sin\theta \rangle$, integrated over $\theta$ (see
Ref.~\cite{COMPASS:2012bfl}).  In the COMPASS acceptance, $\theta$ peaks near
$\pi/2$ with $\langle \sin\theta \rangle = 0.94$, and the $\cos\theta$
distribution is symmetric about zero. The fit function includes the $\sin\phiRS$
modulation, and the extracted fitted amplitudes are normalized by
$\PT\,f(x,y)\,\Dnn$. The average target polarization during the 2022 data-taking
was approximately $0.50$ with a relative uncertainty of about 3\%, estimated
from regular NMR calibrations. The dilution factor varies between 0.35 and 0.45
depending on kinematics. Its uncertainty is estimated to be 2\%. Systematic
uncertainties are assessed by varying the event selection criteria, testing
alternative binning and fit ranges, and evaluating the stability of the results
over the different data-taking periods. In addition, so-called false asymmetries
are extracted by reassigning the target polarization orientation in cells and
found to be consistent with zero within statistical uncertainties. All these
studies confirm the robustness of the extraction method. Further details of the
analysis technique can be found in Refs.~\cite{COMPASS:2012bfl,COMPASS:2014ysd}.

\section{Results}

\begin{figure*}[t]
\centering
\includegraphics[width=0.9\textwidth]{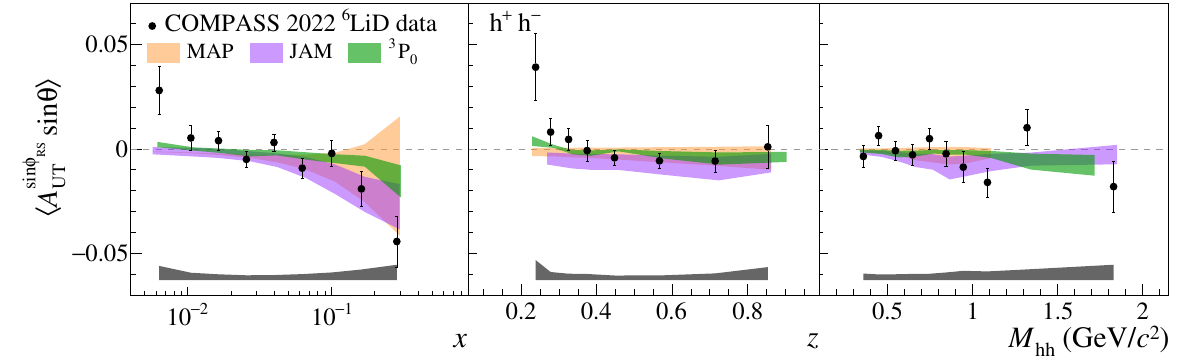}
\caption{Measured dihadron TSA as a function of $\xbj$, $\zh$, and $\Mhh$. The
  gray bands illustrate the associated systematic uncertainties. For comparison,
  the calculations from different models based on Refs.~\cite{Cocuzza:2023vqs}
  (JAM),~\cite{Radici:2018iag} (MAP) and ~\cite{Kerbizi:2023cde} ($^3P_0$) are
  also shown.}
\label{fig:Col_theor}
\end{figure*}
The extracted dihadron TSAs are presented in Fig.~\ref{fig:Col_theor} as a
function 
$\xbj$, $\zh$, and $\Mhh$. 
The statistical uncertainties on the extracted asymmetries range from 0.5\% to
2\% depending on the kinematic bin, while the total point-to-point systematic
uncertainties, shown by the gray bands, are below 60\% of the statistical
ones. The uncertainties are reduced by a factor of about 2-3 compared to the
previous COMPASS deuteron measurement~\cite{COMPASS:2012bfl}.  No significant
signal is observed over the full kinematic range. The measured asymmetries
remain small, typically below the percent level, and are statistically
compatible with zero in most bins. This behavior is qualitatively consistent
with expectations for a deuteron target, where cancellations between $u$- and
$d$-quark contributions suppress the net asymmetry. A hint of a nonzero
asymmetry is observed at large $\xbj$, where the transversity distribution is
expected to be sizable and the $u$-quark contribution dominates. No significant
signal is observed in the $\rho^0$ resonance region.

The overall trend of the asymmetries agrees with the JAM (orange band) and MAP
(purple band) global fit predictions~\cite{Radici:2018iag,Cocuzza:2023vqs},
which are based on $e^{+}e^{-}$, SIDIS, and $pp$ two-hadron data TSA, including,
in particular, previous COMPASS deuteron and proton
results~\cite{COMPASS:2012bfl,COMPASS:2014ysd}.  The dihadron FFs and $h_1^{u}$
and $h_1^{d}$ are extracted simultaneously in the global QCD fits. The
difference between the two models and the size of the shown statistical
uncertainty bands is driven by the differences in the approaches, the data sets
used, and various constraints and assumptions adopted by the groups.

In Fig.~\ref{fig:Col_theor}, the measured asymmetries are also compared with
results from simulations of DIS events on a transversely polarized deuteron
target using the string+$^3P_0$ model (green
band)~\cite{Kerbizi:2018qpp,Kerbizi:2021gos}. In this approach, the color field
between the scattered quark and the target remnant is modeled by a relativistic
string.  The breaking of the string into smaller segments describes the
hadronization process. It occurs through tunneling of quark–antiquark pairs in
the relative $^3P_0$ state, thus correlating the spin and transverse momentum of
the quarks~\cite{Artru:2010st}. Such mechanism gives rise to a nonzero Collins
FF and $\iFF$ if the fragmenting quark is transversely
polarized~\cite{Kerbizi:2018qpp}.
The simulations were performed with the PYTHIA generator~\cite{Bierlich:2022pfr}
using the StringSpinner package~\cite{Kerbizi:2023cde}. The latter enables spin
effects via the string+$^3P_0$ model. The polarization of the fragmenting quark
is parameterised using the point-by-point extractions of $h_1^{u_v}(x)$ and
$h_1^{d_v}(x)$ obtained in Ref.~\cite{COMPASS:2023vhr}.  As seen in the figure,
the simulated asymmetries agree with the data within the uncertainties across
all kinematic variables.  The error bands reflect the statistical uncertainties
of the parameterised transversity PDFs, while statistical uncertainties of the
simulations are negligible.

\begin{figure}[tbh]
\centering
\includegraphics[width=0.50\textwidth]{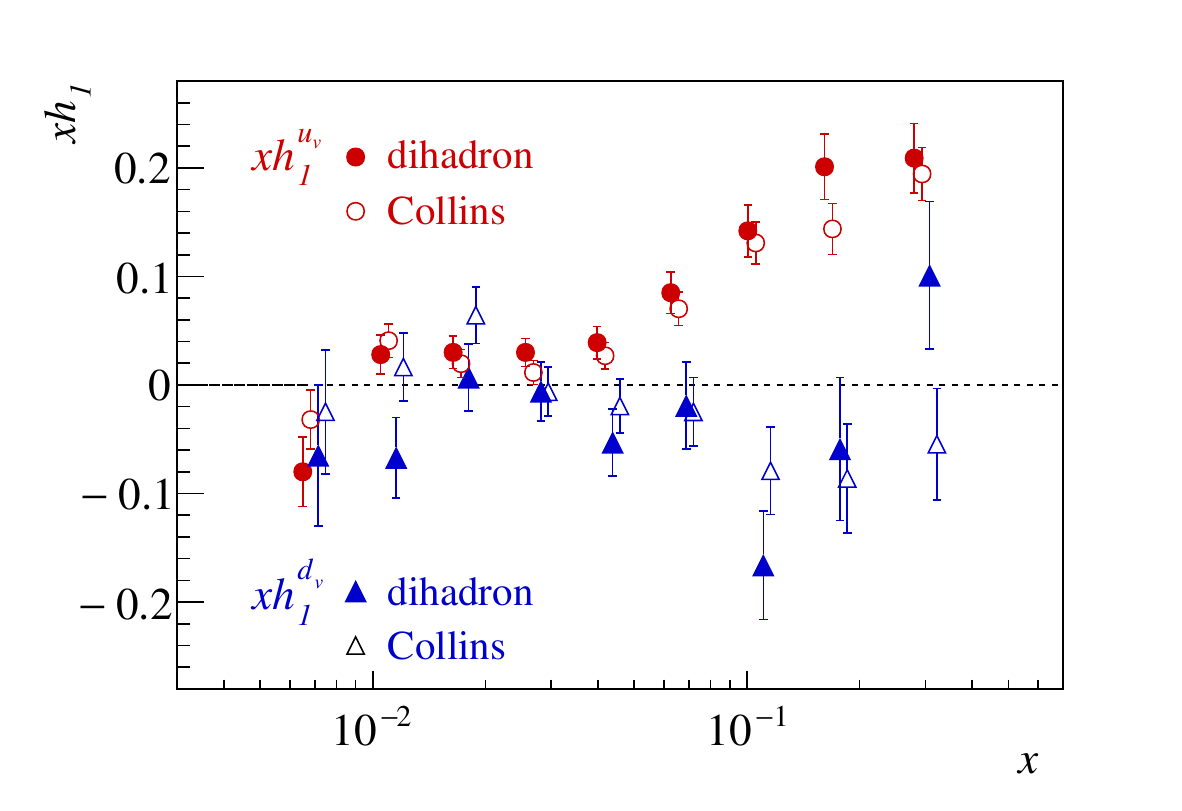}
\caption{The $u_v$ (red circle) and $d_v$ (blue triangles) valence-quark
  transversity in the different $x$ bins, extracted from the full set of COMPASS
  measurements of the dihadron TSAs (closed points) and the Collins asymmetries
  (open points)~\cite{COMPASS:2023vhr}. Error bars represent statistical
  uncertainties.}
\label{fig:h12hc}
\end{figure}

The improvement in the knowledge of the transversity distributions achieved with
the present measurement has been evaluated by extracting, point by point,
$h_1^{u_v}(x)$ and $h_1^{d_v}(x)$ from the dihadron TSAs, following the
leading-order procedure of Ref.~\cite{Martin:2014wua}.  This extraction method
requires neither parameterizations of transversity nor of the fragmentation
functions, and it does not rely on Monte Carlo simulations. As in the case of
the Collins asymmetries~\cite{COMPASS:2023vhr}, the essential inputs are the
measurements of the TSAs in SIDIS off proton and deuteron targets in the same
kinematic range, and the corresponding asymmetries measured in $e^+e^-$
annihilation.

Both $h_1^{u_v}$ and $h_1^{d_v}$ have been extracted from the dihadron TSAs
using, first, only the previously published COMPASS
results~\cite{COMPASS:2012bfl,COMPASS:2014ysd} and, subsequently, the full data
set including the present high-statistics deuteron measurement.  With the
complete set of COMPASS results, the statistical uncertainties are reduced by
about a factor of two for both $h_1^{u_v}$ and $h_1^{d_v}$, clearly illustrating
the impact of the new measurement.  The results obtained using the full COMPASS
data are shown in Fig.~\ref{fig:h12hc} as closed red circles and blue triangles.
The $u$-quark transversity is well determined and clearly positive in the
valence region, while $h_1^{d_v}$ is predominantly negative and lies within the
Soffer bound~\cite{Soffer:1994ww}, which is not imposed in the extraction.  For
comparison, the open points in Fig.~\ref{fig:h12hc} show the transversity values
obtained from the point-by-point extraction~\cite{COMPASS:2023vhr} using the
measured Collins asymmetries.  The agreement between the valence transversities
extracted from the Collins TSAs, based on transverse-momentum–dependent
factorization of single-hadron production, and from the dihadron TSAs, based on
collinear factorization of dihadron production, is quite good. This consistency
supports the validity of both frameworks, in line with the conclusion drawn from
the COMPASS study of the transverse-momentum-weighted Sivers
asymmetries~\cite{COMPASS:2018ofp}.

\section{Conclusions}

Transverse-spin-dependent dihadron asymmetries have been measured using the 2022
COMPASS data collected with a transversely polarised deuteron target.  The
statistical precision of the present results is comparable with that of the
existing proton data and represents a substantial improvement over the previous
measurement from the 2002–2004 deuteron data.  This enables a more detailed
investigation of the transverse spin structure of the nucleon within the
collinear framework.  The results show a non-zero trend at large $\xbj$,
consistent with expectations for the transversity distribution, while no
significant signal is observed across the rest of the kinematic range. A
comparison with recent theoretical predictions based on global fits reveals good
agreement.  The new data has an important potential to reduce the systematic and
statistical uncertainties and constrain the parameterizations for the
transversity PDFs. A bin-by-bin extraction of the transversity PDFs demonstrates
that the deuteron data provide crucial sensitivity, especially to the $d$-quark
transversity distribution, complementing existing measurements on proton targets
and contributing to a more balanced flavour separation.

\section{Acknowledgments}

We gratefully acknowledge the support of the CERN management and staff and the
skill and effort of the technicians of our collaborating institutes.  We also
thank JAM and MAP collaborations, in particular D. Pitonyak, C. Cocuzza and
M. Radici for providing us with their calculations.

\bibliography{COMPASS_2h_2025_biblio}{}
\clearpage

\end{document}

%% file: Authors_list_2025.11.22_PRL_D_SIDIS_2h_TSAs.tex
\center{\textbf{The COMPASS Collaboration}}

\vspace{10pt}
\begin{flushleft}
G.~D.~Alexeev$^\textrm{{\footnotesize\hyperlink{hl:dubna}{28}}}$\orcidlink{0009-0007-0196-8178},
M.~G.~Alexeev$^\textrm{{\footnotesize\hyperlink{hl:turin_u}{20},\hyperlink{hl:turin_i}{19}}}$\orcidlink{0000-0002-7306-8255},
C.~Alice$^\textrm{{\footnotesize\hyperlink{hl:turin_u}{20},\hyperlink{hl:turin_i}{19}}}$\orcidlink{0000-0001-6297-9857},
A.~Amoroso$^\textrm{{\footnotesize\hyperlink{hl:turin_u}{20},\hyperlink{hl:turin_i}{19}}}$\orcidlink{0000-0002-3095-8610},
V.~Andrieux$^\textrm{{\footnotesize\hyperlink{hl:illinois}{33}}}$\orcidlink{0000-0001-9957-9910},
V.~Anosov$^\textrm{{\footnotesize\hyperlink{hl:dubna}{28}}}$\orcidlink{0009-0003-3595-9561},
S.~Asatryan$^\textrm{{\footnotesize\hyperlink{hl:aanl}{1}}}$\orcidlink{0009-0005-2672-8707},
K.~Augsten$^\textrm{{\footnotesize\hyperlink{hl:praguectu}{4}}}$\orcidlink{0000-0001-8324-0576},
W.~Augustyniak$^\textrm{{\footnotesize\hyperlink{hl:warsaw}{23}}}$,
C.~D.~R.~Azevedo$^\textrm{{\footnotesize\hyperlink{hl:aveiro}{26}}}$\orcidlink{0000-0002-0012-9918},
B.~Badelek$^\textrm{{\footnotesize\hyperlink{hl:warsawu}{25}}}$\orcidlink{0000-0002-4082-1466},
R.~Beck$^\textrm{{\footnotesize\hyperlink{hl:bonniskp}{8}}}$,
J.~Beckers$^\textrm{{\footnotesize\hyperlink{hl:munichtu}{12}}}$\orcidlink{0009-0009-7186-255X},
Y.~Bedfer$^\textrm{{\footnotesize\hyperlink{hl:saclay}{6}}}$\orcidlink{0000-0002-5198-1852},
V.~Benesova$^\textrm{{\footnotesize\hyperlink{hl:praguecu}{5}}}$\orcidlink{0009-0003-4051-1542},
J.~Bernhard$^\textrm{{\footnotesize\hyperlink{hl:cern}{30}}}$\orcidlink{0000-0001-9256-971X},
F.~Bradamante$^\textrm{{\footnotesize\hyperlink{hl:triest_i}{17}}}$\orcidlink{0000-0001-6136-376X},
A.~Bressan$^\textrm{{\footnotesize\hyperlink{hl:triest_u}{18},\hyperlink{hl:triest_i}{17},\hyperlink{hl:*}{*}}}$\orcidlink{0000-0002-3718-6377},
W.-C.~Chang$^\textrm{{\footnotesize\hyperlink{hl:taipei}{31}}}$\orcidlink{0000-0002-1695-7830},
C.~Chatterjee$^\textrm{{\footnotesize\hyperlink{hl:triest_i}{17},\hyperlink{hl:a}{a}}}$\orcidlink{0000-0001-7784-3792},
M.~Chiosso$^\textrm{{\footnotesize\hyperlink{hl:turin_u}{20},\hyperlink{hl:turin_i}{19}}}$\orcidlink{0000-0001-6994-8551},
S.-U.~Chung$^\textrm{{\footnotesize\hyperlink{hl:munichtu}{12},\hyperlink{hl:k}{k},\hyperlink{hl:k1}{k1}}}$,
A.~Cicuttin$^\textrm{{\footnotesize\hyperlink{hl:triest_i}{17},\hyperlink{hl:triest_a}{16}}}$\orcidlink{0000-0002-3645-9791},
M.~L.~Crespo$^\textrm{{\footnotesize\hyperlink{hl:triest_i}{17},\hyperlink{hl:triest_a}{16}}}$\orcidlink{0000-0002-5483-3388}
% ...

\vspace{10pt}
\hypertarget{hl:aanl}{$^\textrm{{\footnotesize 1}}$\footnotesize~A.I. Alikhanyan National Science Laboratory, 2 Alikhanyan Br. Street, 0036, Yerevan, Armenia$^\textrm{{\tiny\hyperlink{hl:A}{A}}}$\\}
\hypertarget{hl:brno}{$^\textrm{{\footnotesize 2}}$\footnotesize~Institute of Scientific Instruments of the CAS, 61264 Brno, Czech Republic$^\textrm{{\tiny\hyperlink{hl:B}{B}}}$\\}
\hypertarget{hl:liberec}{$^\textrm{{\footnotesize 3}}$\footnotesize~Technical University in Liberec, 46117 Liberec, Czech Republic$^\textrm{{\tiny\hyperlink{hl:B}{B}}}$\\}
\hypertarget{hl:praguectu}{$^\textrm{{\footnotesize 4}}$\footnotesize~Czech Technical University in Prague, 16636 Prague, Czech Republic$^\textrm{{\tiny\hyperlink{hl:B}{B}}}$\\}
\hypertarget{hl:praguecu}{$^\textrm{{\footnotesize 5}}$\footnotesize~Charles University, Faculty of Mathematics and Physics, 12116 Prague, Czech Republic$^\textrm{{\tiny\hyperlink{hl:B}{B}}}$\\}
\hypertarget{hl:saclay}{$^\textrm{{\footnotesize 6}}$\footnotesize~IRFU, CEA, Universit\'e Paris-Saclay, 91191 Gif-sur-Yvette, France\\}
\hypertarget{hl:bochum}{$^\textrm{{\footnotesize 7}}$\footnotesize~Universit\"at Bochum, Institut f\"ur Experimentalphysik, 44780 Bochum, Germany$^\textrm{{\tiny\hyperlink{hl:C}{C}}}$\\}
\hypertarget{hl:bonniskp}{$^\textrm{{\footnotesize 8}}$\footnotesize~Universit\"at Bonn, Helmholtz-Institut f\"ur  Strahlen- und Kernphysik, 53115 Bonn, Germany$^\textrm{{\tiny\hyperlink{hl:C}{C}}}$\\}
\hypertarget{hl:bonnpi}{$^\textrm{{\footnotesize 9}}$\footnotesize~Universit\"at Bonn, Physikalisches Institut, 53115 Bonn, Germany$^\textrm{{\tiny\hyperlink{hl:C}{C}}}$\\}
\hypertarget{hl:freiburg}{$^\textrm{{\footnotesize 10}}$\footnotesize~Universit\"at Freiburg, Physikalisches Institut, 79104 Freiburg, Germany$^\textrm{{\tiny\hyperlink{hl:C}{C}}}$\\}
\hypertarget{hl:mainz}{$^\textrm{{\footnotesize 11}}$\footnotesize~Universit\"at Mainz, Institut f\"ur Kernphysik, 55099 Mainz, Germany$^\textrm{{\tiny\hyperlink{hl:C}{C}}}$\\}
\hypertarget{hl:munichtu}{$^\textrm{{\footnotesize 12}}$\footnotesize~Technische Universit\"at M\"unchen, Physik Dept., 85748 Garching, Germany$^\textrm{{\tiny\hyperlink{hl:C}{C}}}$\\}
\hypertarget{hl:munichuni}{$^\textrm{{\footnotesize 13}}$\footnotesize~Ludwig-Maximilians-Universit\"at, 80539 M\"unchen, Germany\\}
\hypertarget{hl:calcutta}{$^\textrm{{\footnotesize 14}}$\footnotesize~Matrivani Institute of Experimental Research \& Education, Calcutta-700 030, India$^\textrm{{\tiny\hyperlink{hl:D}{D}}}$\\}
\hypertarget{hl:telaviv}{$^\textrm{{\footnotesize 15}}$\footnotesize~Tel Aviv University, School of Physics and Astronomy, 69978 Tel Aviv, Israel$^\textrm{{\tiny\hyperlink{hl:E}{E}}}$\\}
\hypertarget{hl:triest_a}{$^\textrm{{\footnotesize 16}}$\footnotesize~Abdus Salam ICTP, 34151 Trieste, Italy\\}
\hypertarget{hl:triest_i}{$^\textrm{{\footnotesize 17}}$\footnotesize~Trieste Section of INFN, 34127 Trieste, Italy\\}
\hypertarget{hl:triest_u}{$^\textrm{{\footnotesize 18}}$\footnotesize~University of Trieste, Dept.\ of Physics, 34127 Trieste, Italy\\}
\hypertarget{hl:turin_i}{$^\textrm{{\footnotesize 19}}$\footnotesize~Torino Section of INFN, 10125 Torino, Italy\\}
\hypertarget{hl:turin_u}{$^\textrm{{\footnotesize 20}}$\footnotesize~University of Torino, Dept.\ of Physics, 10125 Torino, Italy\\}
\hypertarget{hl:miyazaki}{$^\textrm{{\footnotesize 21}}$\footnotesize~University of Miyazaki, Miyazaki 889-2192, Japan$^\textrm{{\tiny\hyperlink{hl:F}{F}}}$\\}
\hypertarget{hl:yamagata}{$^\textrm{{\footnotesize 22}}$\footnotesize~Yamagata University, Yamagata 992-8510, Japan$^\textrm{{\tiny\hyperlink{hl:F}{F}}}$\\}
\hypertarget{hl:warsaw}{$^\textrm{{\footnotesize 23}}$\footnotesize~National Centre for Nuclear Research, 02-093 Warsaw, Poland$^\textrm{{\tiny\hyperlink{hl:G}{G}}}$\\}
\hypertarget{hl:warsawtu}{$^\textrm{{\footnotesize 24}}$\footnotesize~Warsaw University of Technology, Institute of Radioelectronics, 00-665 Warsaw, Poland$^\textrm{{\tiny\hyperlink{hl:G}{G}}}$\\}
\hypertarget{hl:warsawu}{$^\textrm{{\footnotesize 25}}$\footnotesize~University of Warsaw, Faculty of Physics, 02-093 Warsaw, Poland$^\textrm{{\tiny\hyperlink{hl:G}{G}}}$\\}
\hypertarget{hl:aveiro}{$^\textrm{{\footnotesize 26}}$\footnotesize~University of Aveiro, I3N, Dept. of Physics, 3810-193 Aveiro, Portugal$^\textrm{{\tiny\hyperlink{hl:H}{H}}}$\\}
\hypertarget{hl:lisbon}{$^\textrm{{\footnotesize 27}}$\footnotesize~LIP, 1649-003 Lisbon, Portugal$^\textrm{{\tiny\hyperlink{hl:H}{H}}}$\\}
\hypertarget{hl:dubna}{$^\textrm{{\footnotesize 28}}$\footnotesize~Affiliated with an international laboratory covered by a cooperation agreement with CERN\\}
\hypertarget{hl:russia}{$^\textrm{{\footnotesize 29}}$\footnotesize~Affiliated with an institute formerly covered by a cooperation agreement with CERN\\}
\hypertarget{hl:cern}{$^\textrm{{\footnotesize 30}}$\footnotesize~CERN, 1211 Geneva 23, Switzerland\\}
\hypertarget{hl:taipei}{$^\textrm{{\footnotesize 31}}$\footnotesize~Academia Sinica, Institute of Physics, Taipei 11529, Taiwan$^\textrm{{\tiny\hyperlink{hl:I}{I}}}$\\}
\hypertarget{hl:taipeincu}{$^\textrm{{\footnotesize 32}}$\footnotesize~Center for High Energy and High Field Physics and Dept.\ of Physics, National Central University, 300 Zhongda Rd., Zhongli 320317, Taiwan$^\textrm{{\tiny\hyperlink{hl:I}{I}}}$\\}
\hypertarget{hl:illinois}{$^\textrm{{\footnotesize 33}}$\footnotesize~University of Illinois at Urbana-Champaign, Dept.\ of Physics, Urbana, IL 61801-3080, USA$^\textrm{{\tiny\hyperlink{hl:J}{J}}}$\\}

\vspace{10pt}
\hypertarget{hl:*}{$^\textrm{{\footnotesize *}}$\footnotesize~Corresponding author\\}
\hypertarget{hl:a}{$^\textrm{{\footnotesize a}}$\footnotesize~Supported by the European Union’s Horizon 2020 research and innovation programme under grant agreement STRONG–2020 - No 824093\\}
\hypertarget{hl:b}{$^\textrm{{\footnotesize b}}$\footnotesize~Retired from Ludwig-Maximilians-Universit\"at, 80539 M\"unchen, Germany\\}
\hypertarget{hl:b1}{$^\textrm{{\footnotesize b1}}$\footnotesize~Supported by the DFG cluster of excellence `Origin and Structure of the Universe' (www.universe-cluster.de) (Germany)\\}
\hypertarget{hl:c}{$^\textrm{{\footnotesize c}}$\footnotesize~Also at ORIGINS Excellence Cluster, 85748 Garching, Germany\\}
\hypertarget{hl:d}{$^\textrm{{\footnotesize d}}$\footnotesize~Also at Institut f\"ur Theoretische Physik, Universit\"at T\"ubingen, 72076 T\"ubingen, Germany\\}
\hypertarget{hl:e}{$^\textrm{{\footnotesize e}}$\footnotesize~Supported by the Max Planck Institute for Physics, 85748 Garching, Germany\\}
\hypertarget{hl:f}{$^\textrm{{\footnotesize f}}$\footnotesize~Present address: NISER, Centre for Medical and Radiation Physics, Bubaneswar, India\\}
\hypertarget{hl:g}{$^\textrm{{\footnotesize g}}$\footnotesize~Also at University of Eastern Piedmont, 15100 Alessandria, Italy\\}
\hypertarget{hl:g1}{$^\textrm{{\footnotesize g1}}$\footnotesize~Supported by the Funds for Research 2019-22 of the University of Eastern Piedmont\\}
\hypertarget{hl:h}{$^\textrm{{\footnotesize h}}$\footnotesize~Also at INFN TIFPA, 38123 Trento, Italy\\}
\hypertarget{hl:h1}{$^\textrm{{\footnotesize h1}}$\footnotesize~Also at Università di Trento, 38123 Trento, Italy\\}
\hypertarget{hl:i}{$^\textrm{{\footnotesize i}}$\footnotesize~Also at Chubu University, Kasugai, Aichi 487-8501, Japan\\}
\hypertarget{hl:j}{$^\textrm{{\footnotesize j}}$\footnotesize~Also at KEK, 1-1 Oho, Tsukuba, Ibaraki 305-0801, Japan\\}
\hypertarget{hl:k}{$^\textrm{{\footnotesize k}}$\footnotesize~Also at Dept.\ of Physics, Pusan National University, Busan 609-735, Republic of Korea\\}
\hypertarget{hl:k1}{$^\textrm{{\footnotesize k1}}$\footnotesize~Also at Physics Dept., Brookhaven National Laboratory, Upton, NY 11973, USA\\}
\hypertarget{hl:l}{$^\textrm{{\footnotesize l}}$\footnotesize~Also at Fairmont State University, Department of Natural Sciences, 1201 Locust Ave, Fairmont, West Virginia 26554, USA\\}
\hypertarget{hl:m}{$^\textrm{{\footnotesize m}}$\footnotesize~Also at Dept.\ of Physics, National Kaohsiung Normal University, Kaohsiung County 824, Taiwan\\}
\hypertarget{hl:n}{$^\textrm{{\footnotesize n}}$\footnotesize~Also at RIKEN Nishina Center for Accelerator-Based Science, Wako, Saitama 351-0198, Japan\\}
\hypertarget{hl:$\dagger$}{$^\textrm{{\footnotesize $\dagger$}}$\footnotesize~Deceased\\}

\vspace{10pt}
\hypertarget{hl:A}{$^\textrm{{\tiny A}}$\footnotesize~Supported by the Higher Education and Science Committee of the Republic of Armenia (Armenia),  within the framework of Scientific Project No 21AG-1C028\\}
\hypertarget{hl:B}{$^\textrm{{\tiny B}}$\footnotesize~Supported by MEYS, Grants LM2023040, LM2018104, LTT17018 and GAUK60121, CZ.02.01.01/00/22\_008/0004632 "FORTE", co-funded by the EU and Charles University Grant PRIMUS/22/SCI/017 (Czech Republic)\\}
\hypertarget{hl:C}{$^\textrm{{\tiny C}}$\footnotesize~Supported by BMBF - Bundesministerium f\"ur Bildung und Forschung (Germany)\\}
\hypertarget{hl:D}{$^\textrm{{\tiny D}}$\footnotesize~Supported by B. Sen fund (India)\\}
\hypertarget{hl:E}{$^\textrm{{\tiny E}}$\footnotesize~Supported by the Israel Academy of Sciences and Humanities (Israel)\\}
\hypertarget{hl:F}{$^\textrm{{\tiny F}}$\footnotesize~Supported by MEXT and JSPS, Grants 18002006, 20540299, 18540281 and 26247032, the Daiko and Yamada Foundations (Japan)\\}
\hypertarget{hl:G}{$^\textrm{{\tiny G}}$\footnotesize~Supported by NCN, Grant 2020/37/B/ST2/01547 (Poland)\\}
\hypertarget{hl:H}{$^\textrm{{\tiny H}}$\footnotesize~Supported by FCT, Grants DOI 10.54499/CERN/FIS-PAR/0022/2019 and DOI 10.54499/CERN/FIS-PAR/0016/2021 (Portugal)\\}
\hypertarget{hl:I}{$^\textrm{{\tiny I}}$\footnotesize~Supported by the Ministry of Science and Technology (Taiwan)\\}
\hypertarget{hl:J}{$^\textrm{{\tiny J}}$\footnotesize~Supported by the National Science Foundation, Grant no. PHY-1506416 (USA)\\}

\end{flushleft}